# Single-cycle optical control of beam electrons

Yuya Morimoto[1,*,+] and Peter Baum[1,2,*]

[1]Ludwig-Maximilians-Universität München, Am Coulombwall 1, 85748 Garching, Germany
[2]University of Konstanz, Universitätsstraße 10, 78457 Konstanz, Germany

*Corresponding to: peter.baum@uni-konstanz.de, yuya.morimoto@fau.de
+Current address: Friedrich-Alexander-Universität Erlangen-Nürnberg, Staudtstraße 1, 91058 Erlangen, Germany

*Dated: April 23, 2020*

**Single-cycle optical pulses with a controlled electromagnetic waveform allow to steer the motion of low-energy electrons in atoms[1], molecules[2], nanostructures[3–6] or condensed-matter[7–9] on attosecond dimensions in time. However, high-energy electrons under single-cycle light control would be an enabling technology for beam-based attosecond physics with free-electron lasers or electron microscopy. Here we report the control of freely propagating keV electrons with an isolated optical cycle of infrared light and create a modulated electron current with a peak-cycle-specific sub-femtosecond structure in time. The evident effects of the carrier-envelope phase, amplitude and dispersion of the optical waveform on the temporal composition, pulse durations and chirp of the free-space electron wavefunction demonstrate the sub-cycle nature of our control. These results create novel opportunities in laser-driven particle acceleration[10,11], seeded free-electron lasers[12], attosecond space-time imaging[13], electron quantum optics[14] and wherever else high-energy electrons are needed with the temporal structure of single-cycle light.**

Modern attosecond science aims at the exploration of ultrafast charge carrier dynamics in complex materials by investigating the time-dependent responses of bound electrons to a single cycle of optical excitation. The electric field as a driving force can for example produce macroscopic electric currents[7,8] or spin waves[9] as a consequence of electronic motion on atomic dimensions. While such light-cycle control of low-energy electrons is ideal for ultrafast electronic operations or for the generation of intense attosecond light pulses, a single-cycle control of the temporal, spatial and energetic structure of high-energy electrons beams would be crucial for laser-driven particle accelerators[10,11], ultrafast electron imaging[15,16], electron-based quantum information technology[17] or attosecond science with free-electron lasers[12]. However, so far only radio-frequency fields[18–20], terahertz radiation[21–24] or optical multi-cycle pulses[10,11,13,25–27] have been employed for the acceleration, compression or metrology of free-space electron pulses on time scales of tens of femtoseconds[19–21,23,24] or in form of multi-pulse sequences[10,11,13,25–27]. Although pioneering electron acceleration experiments were reported with few-cycle laser pulses[28–30], it remains to be established whether and how an isolated optical field cycle can control the temporal shape of free-space electrons at keV to MeV energies with attosecond precision, in order to merge the unprecedented power and brightness of high-energy electron beams with the temporal structure provided by single-cycle laser light.

  The sketch and concept of our experiment is shown in Fig. 1a. A femtosecond Yb:KGW laser (magenta) is used for generating a beam of 70-keV electron pulses (blue) and for pumping a mid-infrared optical parametric amplifier[31] for single-cycle waveform generation. Briefly, near-infrared pulses (orange) from the optical parametric amplifier (NOPA) are mixed with fundamental pulses from the Yb:KGW laser (magenta) in a $LiGaS_2$ crystal for difference-frequency generation. Passive phase locking produces mid-infrared pulses with a stable carrier-envelope phase. The spectrum (see Figure 1b) spans from 26 THz (11.5 μm) to 62 THz (4.8 μm) at −20 dB level, exceeding one optical octave. The temporal waveform is characterized by electro-optical sampling and shown in Fig. 1c. The main field cycle at $t \approx 0$ fs is 1.8 times stronger than the adjacent positive peaks at $t \approx \pm 23$ fs and the pulse shape therefore allows a sub-cycle control of the electron beam. A split and displaced parabolic mirror (yellow) focuses two such pulses onto ultrathin membranes (green) for temporal modulation of the electron wavepacket (blue) and its subsequent streaking characterization. The amplitude and the carrier-envelope phase (CEP) of the two mid-infrared pulses are adjusted independently (see Fig. 1a).

In order to control the electron beam by a single cycle of light (red), we invoke electron-transparent metallic membranes with an extremely broad bandwidth as the modulation elements for photon-electron energy exchange[32,33]. Free-standing silicon nitride membranes are coated with ~10 nm of aluminium. Although this coating is >500 times thinner than a wavelength, we find that the membranes reflect our single-cycle mid-infrared pulses efficiently over the full range of their octave-broad spectrum. The 70-keV electrons pass though the membrane within a time of <140 as and are therefore injected into the optical electromagnetic field on the backside within sub-cycle time. This abrupt injection from a field-free region before the membrane into the optical field cycles on the backside causes a time-dependent electron energy modulation that follows the temporal integral of the optical waveform[32,34]. At the first membrane, the one for single-cycle control, we apply an optical peak field strength of ~25 MV/m. Vacuum is a dispersive medium for our electrons and the cycle-induced energy modulation is therefore transformed by propagation into a modulation of current density in time[13]. We characterize the final electron pulse shape at a distance of ~12 mm at a second metal membrane under illumination of a stronger single-cycle field at ~200 MV/m that produces a field-driven sideways electron deflection as a function of time[21]. This real-space streaking by a single-cycle field isolates the compressed electron pulses in transverse momentum space with sub-femtosecond precision and therefore provides a direct metrology of the effects of the first interaction's single-cycle control in time.

Figure 1d shows the observed streaking pattern on the screen, plotted as a function of the delay $\Delta t$ between the control field and the streaking field. The largest streaking angle is more than 0.4 mrad, corresponding to the absorption or emission of more than 900 photons at 6.9 μm central wavelength. The highest streaking speed around $\Delta t \approx \pm 11$ fs is ±0.1 mrad/fs, enabling attosecond time resolution[13,34]. There are pronounced streaking oscillations of sideways deflection as a function of delay, but unlike in previous cases with multi-cycle fields[13] we see here a streaking signal that does not repeat itself before or after one optical cycle of delay. In other words, the streaking peak shapes around $\Delta t \approx 0$ fs (peak 0; dotted rectangle) and around $\Delta t \approx \pm 23$ fs (peaks +2 and -2) differ substantially in magnitude and shape. These observations suggest the presence of one exceptional peak of electron density with sub-cycle duration within the pattern of compressed electron density in our beam.

The left panel of Fig. 1e shows a magnified view of the large-angle streaking signal around $\Delta t \approx 0$ fs (see dotted rectangle in Fig. 1d). The measured streaking intensity shows a broad maximum at the turning point around ~0.42 mrad, a hole with two separate borders in time at ~0.35 mrad (dotted line) and a temporally washed-out pattern at lower angles. Results of a numerical simulation (see Methods) are depicted in the right panel of Fig. 1e for an electron pulse duration of 1.0 fs (full-width-at-half-maximum). A more detailed comparison between experiment and simulations is shown in Fig. 1f, where we plot a cut through Fig. 1e at a deflection angle of 0.37 mrad. We see a high streaking intensity at $\Delta t \approx -2$ fs, followed by a minimum at $\Delta t \approx 0$ fs and again a maximum at $\Delta t \approx +2$ fs. This double peak in time with dip in the middle resembles almost the classical time-dependent deflection dynamics that would occur for electron pulses of negligible duration in time[21]. Comparison of the measured data in Fig. 1f (black dots) to the results of the simulations (blue lines) indicates an electron pulse duration somewhere between 0.5 fs (dashed blue line) and 1.0 fs (solid blue line).

For a more profound characterization and for understanding the role of potential electron density peaks from adjacent compression cycles (see Fig. 2a), we consider the single-cycle nature of the streaking deflection at high angles (see Fig. 1c) and apply a numerical deconvolution and fitting procedure (see Methods). Basically, streaking angles above 0.35 mrad can only originate from an isolated but finite range of time around $t$ = 0 with few-femtosecond duration (see Fig. 1c). A numerical consideration of the angular divergence of the electron beam, causing angular blurring in Fig. 1e, provides a temporal instrument response function that we use for deconvolution (see Methods). Figure 2b shows the resulting time-dependent electron current density as a function of the peak field strength of the central control cycle. A peaking electron density emerges at points in time where the electric field has positive peaks (dotted lines). There, we have a close-to-linear time-dependent acceleration; preceding electrons are decelerated and trailing electrons are accelerated. In contrast, there is no electron pulse compression for optical cycles with negative field strengths.

Although the secondary positive field crests (peaks A, B and D in Fig. 2b) can also create a bunched electron current, these cycles have lower field strengths and therefore produce distinguishable compression results. At ~25 MV/m, for example, peak C is well compressed, but all adjacent peaks are still long. A higher compression strength (lower panels) disperses the central peak towards longer duration and instead compresses the adjacent peaks A, B and D. Figure 2c shows the retrieved electron pulse durations (see Methods) as a function of the compression strength; the dashed lines denote the results of quantum mechanical simulations[35]. We see that each of the optical cycles produces a minimum electron pulse duration at substantially different compression strength. Exceptional conditions are for example achieved at ~25 MV/m (third panel of Fig. 2b), where the central electron pulse C is almost isolated in time (assuming a full-width-at half-maximum criterion), or again at ~80 MV/m, where we observe an over-dispersion (depletion) of peaks B-D but emergence of an isolation of peak A (see Fig. S2). In principle, any optical field cycle of the compression waveform that is unique with respect to the other field cycles can selectively optimize the compression of a single peak density of electrons in time, and subsequent single-cycle time-dependent sideways deflection (Fig. 1d) into a high-angle aperture isolates this pulse from the satellites and time-independent background.

Further evidence for the single-cycle control of our electron beam can be obtained from a scan of the carrier-envelope phase $\phi_{CEP}$ of the compression waveform. Figures 2d-e show the raw high-angle streaking data at >0.35 mrad (black dots) and the retrieved electron density (blue) as a function of time for $\phi_{CEP} \approx 0$ (cosine waveform) and $\phi_{CEP} \approx \pm\pi$ (minus-cosine waveform); a continuous scan of $\phi_{CEP}$ is reported in the supplementary materials. With the cosine-shaped modulation field, we create an electron pulse structure with one exceptional peak (upper panels in Figs. 2d-e) but the minus-cosine-like modulation field produces two almost equally high peaks (lower panels in Figs. 2d-e) that originate from the two previously negative field cycles at ±11 fs (see Fig. 2a). Figure 2f depicts the compressed electron pulse durations as a function of the carrier-envelope phase for the three peaks B-D in Fig. 2d. The middle peak (C) has the shortest duration at $\phi_{CEP} = 0$ (cosine field) while peak B becomes shorter and peak D becomes longer when increasing $\phi_{CEP}$. At $\phi_{CEP} \approx \pm\pi$ (minus-cosine field), peaks exchange places (see supplementary materials) and there emerges a double-peak structure (see lower panels in Figs.

2d-e). We conclude that a simple change of the absolute direction of the control field produces electron current densities of substantially different shape in time.

It is also possible to produce a multi-pulse sequence of few-femtosecond electron pulses with longer mid-infrared pulses (see Methods) in which there are no single-cycle effects. At an optical pulse duration of ~800 fs (see Fig. 3a), the field cycles used for modulation and streaking have approximately equal field strength over the entire duration of the incoming electron pulses from the source (~500 fs). The observed delay-dependent streaking signal (Fig. 3b) therefore consists of tens of coherent oscillations over hundreds of femtoseconds, demonstrating the creation of multiple electron pulses with sub-cycle duration in synchrony to the optical cycles of the streaking field[13]. Carrier-envelope phase effects are indistinguishable from a delay in this experiment. Figure 3c shows the evaluated electron pulse duration as an average value over all the individual pulses in the sequence. There is an optimum shortness at ~25 MV/m, similar to the best field strength needed for the single-cycle control (compare Fig. 2b). The shortest average pulse duration of 3.1±0.7 fs (full width at half maximum) is a little longer than the best single-cycle results (≤1 fs; see Figs. 1f and 2c), because inelastic energy losses at the modulation membrane or differences of chirp of the involved laser and electron pulses may contribute to a varying duration of the different individual pulses in the burst. All such effects are irrelevant in case of the single-cycle control. Nevertheless, a cycle-locked few-femtosecond electron pulse train in which the individual pulses are separated by tens of femtoseconds instead of few femtoseconds for near-infrared excitation[10,11,13,25–27] can be useful for waveform electron microscopy[13], electron acceleration[10,11] or quantum control of electron wavepackets[14,36] in cases where isolated attosecond electron pulses are not necessarily required[13].

In summary, the reported results demonstrate that the temporal shape of freely propagating high-energy electrons can be controlled and shaped by the field cycles of single-cycle light. Attosecond science with electrons in isolated electromagnetic field cycles is therefore advanced from eV-level energies[1–9] into the sub-relativistic and relativistic regime of the free-space electron beams. The only prerequisite of our approach is support of at least approximately one optical octave of bandwidth by the modulation element. Sub-cycle optical control of the energy, temporal shape and space-time correlations of beam electrons with attosecond precision will for example enable the injection of isolated attosecond electron pulses into laser-driven particle accelerators[10,11], the attosecond-Angstrom imaging of complex material dynamics with electron microscopy or diffraction[13], the coherent control of quantum systems[37] and radiation processes[38] or the ultrafast modulation and tomography of free-electron quantum states[14,17,26] in energy and time. More generally, unifying the power and brightness of modern free-space electron beams with the ultimate control of time by modern attosecond science may provide a general novel tool for exploring and controlling complex materials at unprecedented levels of flexibility, power, energy and time.

**Acknowledgements.** This work was supported by the European Research Council (CoG No. 647771) and the Munich-Centre for Advanced Photonics. We thank Bo-Han Chen and Christina Hofer for helpful discussions on mid-infrared optics, Simon Stork and Jerzy Szerypo for membrane coating, Alexander Gliserin for help with measurement automation and Ferenc Krausz for generous supply of laboratory infrastructure. YM acknowledges general support from Peter Hommelhoff.

**Author contribution.** YM and PB conceived the experiment. YM performed the experiment and analysed the data. PB and YM discussed the results and wrote the manuscript.

**Competing interests.** The authors declare no competing interests.

**Supplementary information.** Supplementary Figs. S1-S3.

**Data availability.** All relevant data and simulation codes are available from the corresponding authors upon reasonable request.

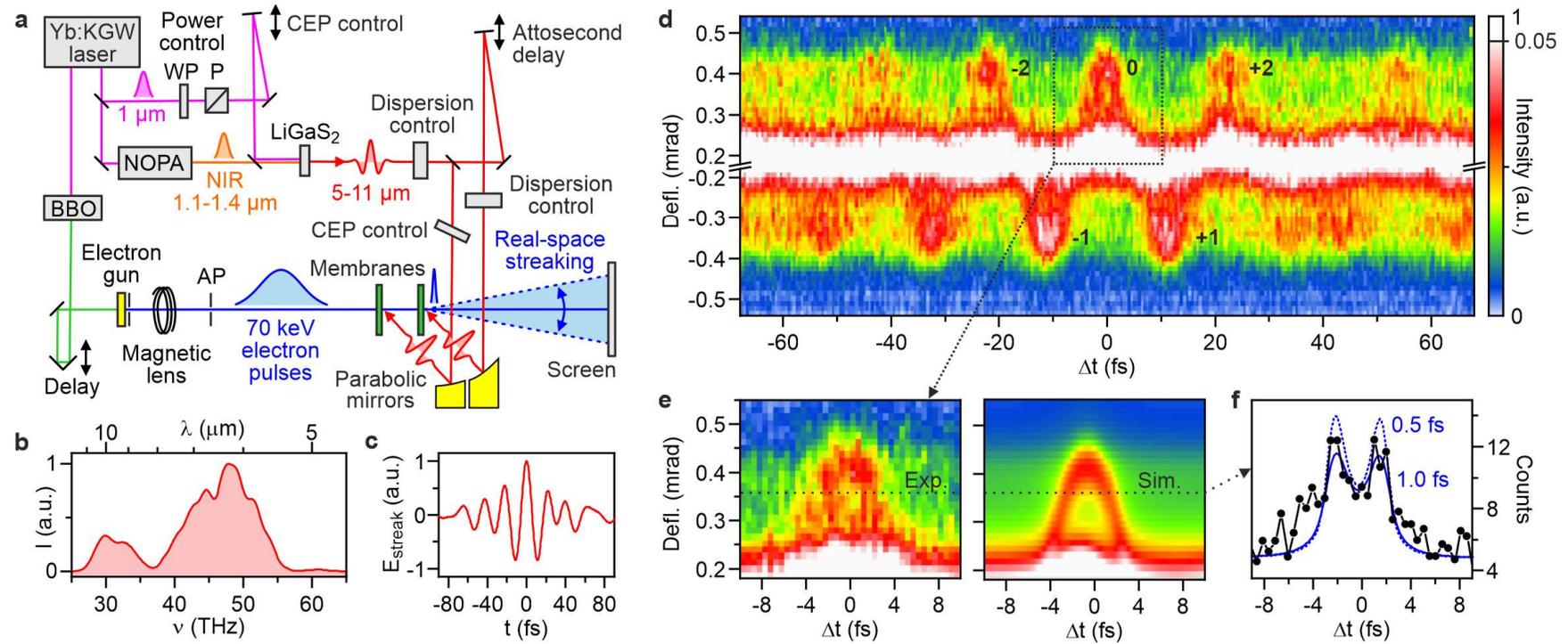

**Figure 1. Single-cycle control of sub-relativistic beam electrons. a.** Concept and experiment. A 70-keV electron beams is temporally modulated and analysed by a single optical cycle of a mid-infrared field (red). WP, waveplate; P, polarizer; AP, aperture. **b.** Spectrum of the mid-infrared pulses. **c.** Electric field waveform of the mid-infrared pulses. **d.** Streaking signal of the temporally modulated electron pulses as a function of the modulation-streaking delay $\Delta t$. **e.** Streaking signal of the central feature (left panel) in comparison to a simulation (right panel) for an electron pulse duration of 1.0 fs (full width at half maximum). **f.** Slice of the streaking pattern at an angle of 0.37 mrad. The observations (black circles) range between a simulation with an electron pule duration of 1.0 fs (blue solid curve) and 0.5 fs (blue dotted curve).

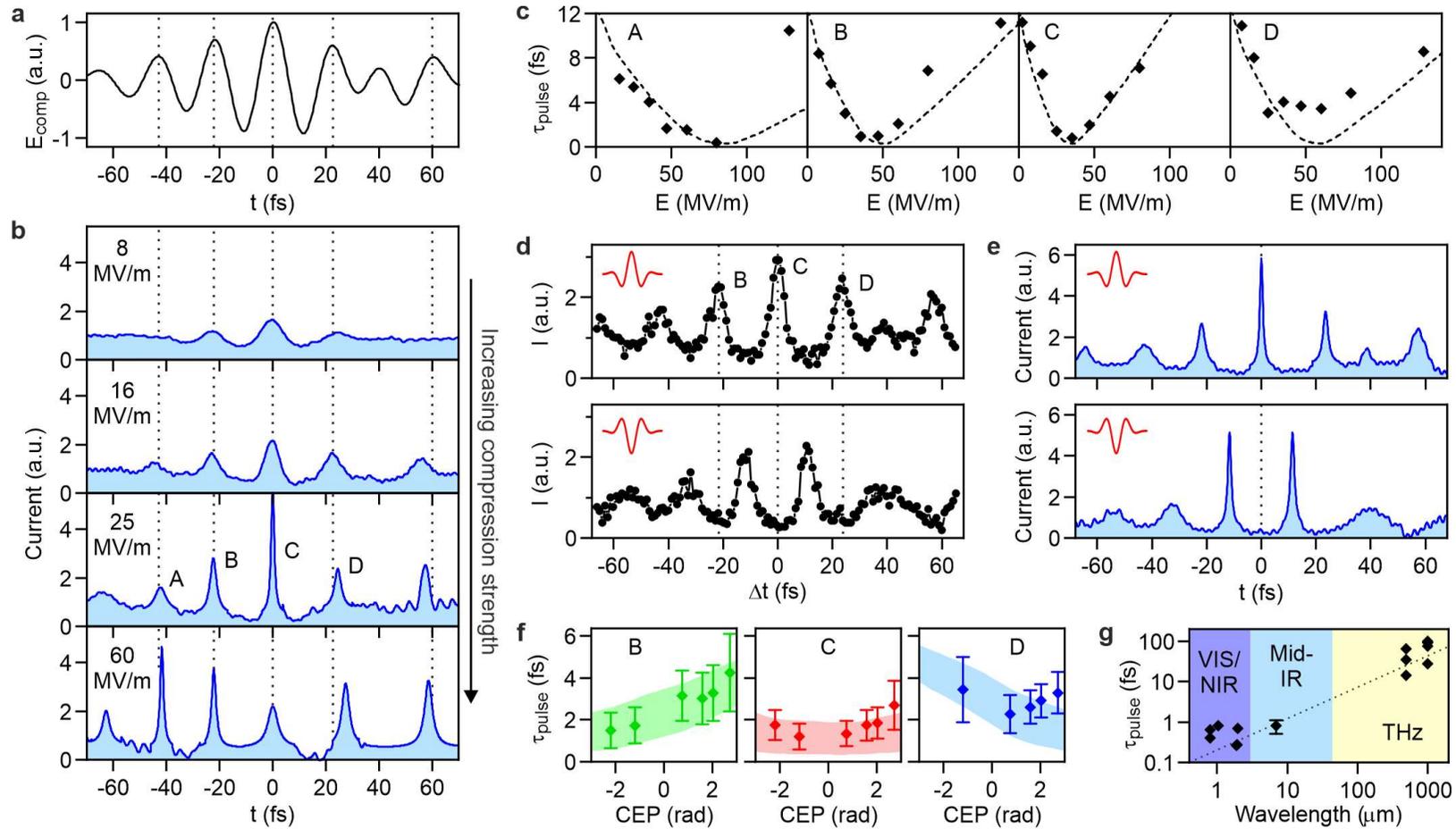

**Figure 2. Electron pulse formation and carrier-envelope phase effects. a.** Waveform of the modulation field. **b.** Electron current density in dependence of an increasing modulation strength. **c.** Electron pulse durations (black squares) for the four peaks A-D (see **b**) as a function of the field strength of the central cycle. The shortest full-width-at-half-maximum durations are 0.6±0.6 fs (A), 0.9±0.8 fs (B), 0.8±0.6 fs (C) and 3.1±1.8 fs (D). Dashed lines depict the result of quantum mechanical simulations. **d-e,** Carrier-envelope phase control. **d.** Time-dependent intensity at highest streaking angle (>0.35 mrad) for a cosine field (upper panel) and for a minus-cosine field (lower panel). **e.** Retrieved electron current density for the two control field shapes. The appearance of a single peak or double peaks is determined by the carrier-envelope phase. **f.** Pulse durations (dots) of the peaks B-D as a function of the carrier-envelope phase. The transparent bands show the results of simulations with error margins. **g.** Survey of the electron pulse durations so far obtained at different control wavelengths. Single-cycle electron control is applicable to any wavelength for which an octave-broad modulation element can be designed (see Methods).

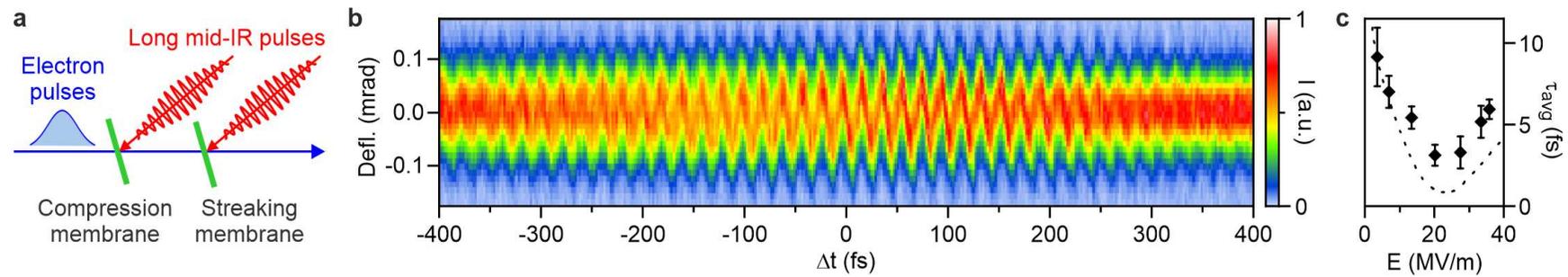

**Figure 3. Free-electron control with multi-cycle mid-infrared fields. a.** Experimental scheme for multi-cycle mid-infrared control. **b.** Measured streaking data. The oscillation period and therefore the temporal separation of the electron pulses are ~21 fs. **c.** Train-averaged electron pulse duration (dots) as a function of the compression field strength in comparison to the results of quantum mechanical simulations (dashed line).